\documentclass[final]{aipproc}

\usepackage{amssymb}
\usepackage{amsmath}
\layoutstyle{8x11double}

\newcommand{\beqn}{\begin{eqnarray}}
\newcommand{\eeqn}{\end{eqnarray}}
\newcommand{\beqs}{\begin{subequations}}
\newcommand{\eeqs}{\end{subequations}}
\newcommand{\eq}[1]{(\ref{#1})}

\newcommand{\ext}{{\mathrm{ext}}}

\begin{document}

\title{Electromagnetically superconducting phase
of QCD vacuum \\ induced by strong magnetic field\footnote{Talk given at ``Confinement IX'', 30 Aug - 3 Sep, 2010, Madrid.}}


\author{M. N. Chernodub\thanks{maxim.chernodub@lmpt.univ-tours.fr; on leave from ITEP, Moscow.}}{
  address={CNRS, Laboratoire de Math\'ematiques et Physique Th\'eorique, Universit\'e Fran\c{c}ois-Rabelais Tours,\\ F\'ed\'eration Denis Poisson, Parc de Grandmont, 37200 Tours, France},
  altaddress={Department of Physics and Astronomy, University of Gent, Krijgslaan 281, S9, B-9000 Gent, Belgium},
  email={maxim.chernodub@lmpt.univ-tours.fr}
}

\begin{abstract}
In this talk we discuss our recent suggestion that the QCD vacuum in a sufficiently strong magnetic field (stronger than $10^{16}$ Tesla) may undergo a spontaneous transition to an {\emph {electromagnetically}} superconducting state. The possible superconducting state is anisotropic (the vacuum exhibits superconductivity only along the axis of the uniform magnetic field) and inhomogeneous (in the transverse directions the vacuum structure shares similarity with the Abrikosov lattice of an ordinary type-II superconductor). The electromagnetic superconductivity of the QCD vacuum is suggested to occur due to emergence of specific quark-antiquark condensates which carry quantum numbers of electrically charged rho mesons. A Lorentz-covariant generalization of the London transport equations for the magnetic-field-induced superconductivity is given.
\end{abstract}

\maketitle

Recently, we suggested that the QCD vacuum in a sufficiently strong magnetic field may undergo a spontaneous transition to an electromagnetically superconducting state~\cite{ref:I}. We stress the word ``\emph{electromagnetic}'' in order to distinguish the proposed superconducting state from the ``color superconductivity'' (which may exist in a sufficiently dense quark matter) and from the ``dual  superconductivity'' associated with confining features of the gluonic fields in the pure Yang-Mills vacuum. We consider QCD at zero temperature.

We suggest that the vacuum becomes electromagnetically superconducting if the strength of the background magnetic field exceeds the critical value 
\beqn
B_c = m_\rho^2/e \approx 10^{16}\,\mbox{Tesla}\,,
\label{eq:eBc}
\eeqn
where $m_\rho = 775.5\,\mbox{MeV}$ is the mass of the $\rho$ meson. 
If the background magnetic field is stronger then the critical value~\eq{eq:eBc}, then the vacuum becomes unstable towards emergence
of the offdiagonal (in flavor space) vector (in coordinate space) quark-antiquark condensates
\beqn
\langle \bar u \gamma_1 d\rangle = \rho(x_\perp)\,,
\qquad
\langle \bar u \gamma_2 d\rangle =  i \rho(x_\perp)\,,
\label{eq:ud:cond}
\eeqn
where $\rho\neq 0$ is the complex scalar field which depends on the transverse coordinates $x_\perp = (x_1,x_2)$,
and the uniform static magnetic field $B$ is set along the $z\equiv x_3$ axis, $\vec B = (0,0,B)$.  
The condensates \eq{eq:ud:cond} carry quantum numbers of the electrically-charged vector $\rho$ mesons with the quark content $\rho^- = d \bar u$ and $\rho^+ = u \bar d$. Thus, the appearance of the condensates \eq{eq:ud:cond} is equivalent, to some extent, to the condensation of the charged $\rho$ mesons. Since the condensed mesons are electrically charged,
their condensation implies, almost automatically, an electromagnetic superconductivity of the condensed state\footnote{
Notice that in a dense isospin--asymmetric matter the longitudinal condensates $\langle d \gamma_{0,3} \bar u\rangle$ may emerge~\cite{ref:isospin} and a superconducting state may also be formed~\cite{Ammon:2008fc}. On the contrary, our condensates~\eq{eq:ud:cond} are spatially-transverse and they appear in the vacuum (with no matter present).}. We argue that the state \eq{eq:ud:cond} is indeed superconducting, and it is accompanied by a superfluidity of the neutral $\rho^{(0)}$ mesons~\cite{ref:I}. 

Let us provide qualitative arguments that the $\rho$-meson condensation is not an absolutely unexpected phenomenon. Consider a simple example of a free relativistic spin-$s$ particle moving in a background of the external magnetic field $B$. The energy levels $\varepsilon$ of the particle are:
\beqn
\varepsilon_{n,s_z}^2(p_z) = p_z^2+(2 n - 2 s_z + 1) |eB| + m^2\,,
\label{eq:energy:levels}
\eeqn
where the integer $n\geqslant 0$ labels the energy levels, and other quantities characterize the properties of the pointlike particle: mass $m$, the projection of the spin $s$ on the field's axis $s_z = -s, \dots, s$, the momentum along the field's axis, $p_z$, and the electric charge $e$.

It is clear from Eq.~\eq{eq:energy:levels} that the ground state corresponds to  $p_z=0$, $n_z = 0$ and $s_z = s$. The ``minimal masses'', corresponding to the ground state energies of the charged pions (with $s=0$) and charged $\rho$ mesons (with $s=1$), are, respectively:
\beqn
m_{\pi^\pm}^2(B) = m_{\pi^\pm}^2 + e B\,, \qquad
m_{\rho^\pm}^2(B) = m_{\rho^\pm}^2 - e B\,.
\label{eq:m2:pi:B}
\eeqn
Thus, the ground state energy of the charged pion is the increasing function of the strength of the magnetic field $B$, while the ground state energy of the charged $\rho$ meson decreases with $B$. When the magnetic field reaches the strength~\eq{eq:eBc} the mass of the charged $\rho$ meson becomes zero. As the field increases further, the ground state energy of the charged $\rho$ mesons becomes purely imaginary thus signaling a tachyonic instability of the QCD ground state. At these magnetic fields the QCD vacuum spontaneously develops the ``$\rho$-meson condensates'' \eq{eq:ud:cond}. Thus, the system becomes electromagnetically superconducting via the condensation of the charged $\rho$ mesons.

The suggested condensation of the $\rho$ mesons is similar to the Nielsen-Olesen instability of the gluonic vacuum in Yang-Mills theory~\cite{ref:NO}, and to the magnetic-field-induced Ambj\o rn--Olesen condensation of the $W$-bosons in the standard electroweak model~\cite{ref:AO}: both the $\rho$ mesons in QCD, the gluons in Yang-Mills theory, and the $W$ bosons in the electroweak model have the anomalously large gyromagnetic ratio, $g=2$ [explicitly implemented in Eq.~\eq{eq:energy:levels}] which supports the mentioned effects. 

Equations~\eq{eq:m2:pi:B} imply, in addition, that in the strong magnetic field the dominant decays of both charged and neutral $\rho$ mesons into pions become kinematically forbidden. The decays will be ``reversed'', and the pions will decay into the $\rho$ mesons~\cite{ref:I}. 

A simple realization of the magnetic-field-induced superconductivity can be found in a quantum electrodynamics for the $\rho$ mesons based on a vector meson dominance model. A relevant part of the Lagrangian is~\cite{ref:QED:rho}:
\beqn
{\cal L} = -\frac{1}{4} \ F_{\mu\nu}F^{\mu\nu}
- \frac{1}{2} (D_{[\mu,} \rho_{\nu]})^\dagger D^{[\mu,} \rho^{\nu]} + m_\rho^2 \ \rho_\mu^\dagger \rho^{\mu}
\nonumber\\
-\frac{1}{4} \ \rho^{(0)}_{\mu\nu} \rho^{(0) \mu\nu}+\frac{m_\rho^2}{2} \ \rho_\mu^{(0)}
\rho^{(0) \mu} +\frac{e}{2 g_s} \ F^{\mu\nu} \rho^{(0)}_{\mu\nu}\,, 
\label{eq:L:rho}
\eeqn
where $D_\mu = \partial_\mu + i g_s \rho^{(0)}_\mu - ie A_\mu$ is the covariant derivative,
$g_s \equiv g_{\rho\pi\pi} \approx 5.88$ is the $\rho\pi\pi$ coupling,
$A_\mu$ is the photon field with the field strength $F_{\mu\nu} = \partial_{[\mu,} A_{\nu]}$, 
$\rho_\mu \equiv \rho^-_\mu= (\rho^{(1)}_\mu - i \rho^{(2)}_\mu)/\sqrt{2}$ and $\rho^{(0)}_\mu \equiv \rho^{(3)}_\mu$
are, respectively, the fields of the charged and neutral vector mesons with the mass~$m_\rho$, 
and $\rho^{(0)}_{\mu\nu} = \partial_{[\mu,} \rho^{(0)}_{\nu]} - i g_s \rho^\dagger_{[\mu,} \rho_{\nu]}$.

The model enjoys the $U(1)$ gauge invariance:
\beqn
U(1)_{\mathrm{e.m.}}: \quad
\left\{
\begin{array}{lcl}
\rho_\mu(x) & \to & e^{i \omega(x)} \rho_\mu(x)\,,\\
A_\mu(x) & \to & A_\mu(x) + \partial_\mu \omega(x)\,. \quad
\end{array}
\right.
\label{eq:gauge:invariance}
\eeqn
The last term in Eq.~\eq{eq:L:rho} describes a nonminimal coupling of the $\rho$ mesons to the electromagnetic field implying
the anomalous gyromagnetic ratio ($g = 2$) of the charged $\rho^\pm$ mesons, and thus playing a crucial role in the emergence of the electromagnetic superconductivity.

\begin{figure}
  \includegraphics[height=.2\textheight]{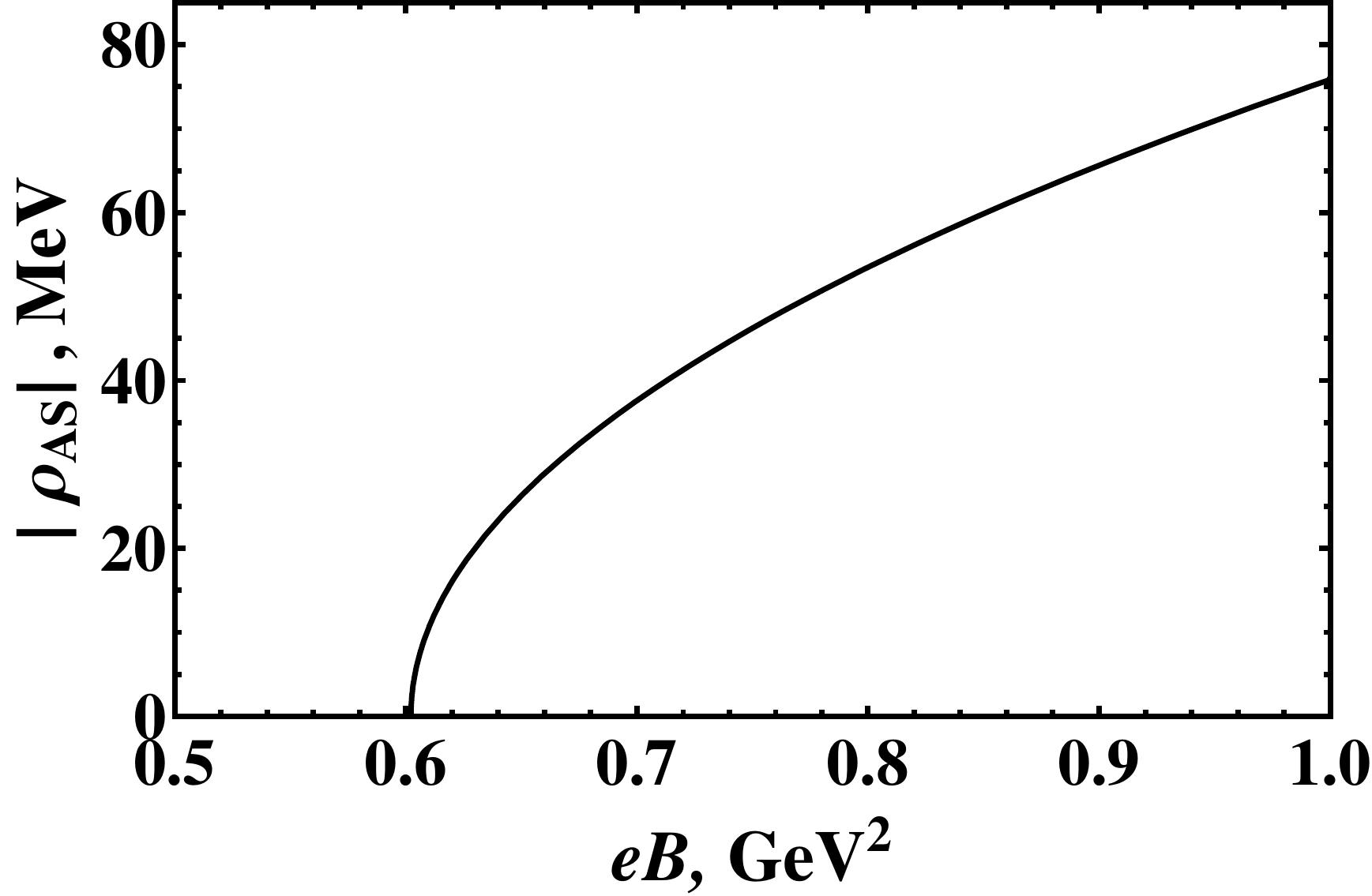}
  \caption{The mean value of the $\rho$ condensate~\eq{eq:ud:cond} as the function of the magnetic field strength $B$ (from Ref.~\cite{ref:I}).}
\label{eq:mean:condensate}
\end{figure}

An explicit solution of the classical equations of motion of the model~\eq{eq:L:rho} in the background of the strong magnetic field shows the spontaneous formation of the 
transversely inhomogeneous condensate~\eq{eq:ud:cond} if the external magnetic field $\vec B \equiv {\vec B}^{\ext}$ exceeds the critical value~\eq{eq:eBc}. The transition is expected to be of the second order with the critical exponent 1/2. In Figure~\ref{eq:mean:condensate} we show the (spatially-averaged) condensate as a function of $B$ .

\begin{figure}
  \includegraphics[height=.17\textheight]{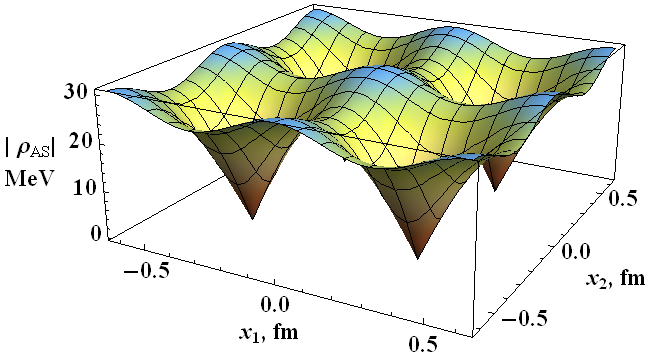}
  \caption{The absolute value of the $\rho$-meson condensate~\eq{eq:ud:cond} in the transverse (with respect to the magnetic field axis) plane at the external field $eB=(800\, \mbox{MeV})^2$. Four unit lattice cells of the $\rho$ vortices are shown (from Ref.~\cite{ref:I}).}
\label{eq:condensate}
\end{figure}

In the transverse (with respect to the magnetic field axis) plane the $\rho$-meson condensate has typical features of the Abrikosov vortex lattice of a mixed state of an ordinary type-II superconductor. It turns out that the $\rho$ condensate supports the new topological object, the ``$\rho$ vortex'', which has typical features of an ordinary Abrikosov vortex~\cite{Abrikosov:1956sx}: in the center $x_0$ of the $\rho$ vortex the condensate vanishes, $\rho(x_0) = 0$, while the phase of the condensate, $\phi = \arg \rho(x)$, has a quantized winding number. The ground state~\eq{eq:ud:cond} is, in fact, the Abrikosov-like lattice made of the $\rho$ vortices (Figure~\ref{eq:condensate}).

The condensate~\eq{eq:ud:cond} ``locks'' rotational and electromagnetic gauge symmetries. Indeed, in the presence of the background magnetic field ${\vec B}$ the group of global rotations of the coordinate space, $SO(3)_{\mathrm{rot}}$, is explicitly broken to its $O(2)_{\mathrm{rot}}$ subgroup generated by rotations around the axis of the magnetic field. Under the global $O(2)_{\mathrm{rot}}$ the field $\rho$ transforms as follows
$\rho(x) \to e^{i \varphi} \rho(x)$,
where $\varphi$ is the azimuthal angle of the rotation around the magnetic axis. Thus, the ground state~\eq{eq:ud:cond}
is invariant under a combination of the global transformation generated by the gauge group~\eq{eq:gauge:invariance} and the
global rotation around the field axis provided the parameters of these transformations are related (``locked'') to each other as follows: $\omega(x) = - \varphi$.  The inhomogeneities in the condensate break the locked subgroup further
down to a discrete subgroup  $G^{\mathrm{lattice}}_{\mathrm{locked}}$ of the rotations of the $\rho$-vortex lattice:
\beqn
U(1)_{\mathrm{e.m.}} \times O(2)_{\mathrm{rot}} \to U(1)_{\mathrm{locked}} \to G_{\mathrm{locked}}^{\mathrm{lattice}}\,.
\label{eq:locking2}
\eeqn

The (super)conducting properties of the system may be studied with the help of the transport equations like the London equations, which -- in a {\it conventional} superconductor -- are usually written as follows:
\beqn
\partial_t \vec j+\vec \nabla q = \lambda^{-2} {\vec E}\,,
\qquad 
\vec \nabla \times \vec j = - \lambda^{-2} {\vec B}\,,
\label{eq:London}
\eeqn
where $\lambda$ is the London penetration depth, $\vec j = \vec j(\vec x,t)$ is the electric current,
$q = q(\vec x,t)$ is the density of electric charge, and $\vec E \equiv {\vec E}^{\ext}$  is the weak ($|\vec E| \ll |\vec B|$) external electric field.
The first equation describes an accelerating flow of the superconducting longitudinal electric current in the presence of the external electric field $\vec E$, while the second equation guarantees the emergence of the circulating transverse electric currents which screen the external magnetic field $\vec B$ inside the superconductor.

The transport equations in the electromagnetically superconducting QCD state are as follows\footnote{An indication of existence of an anisotropically conducting state with $j_z \sim E_z$ was recently found in quenched simulations of lattice QCD at strong magnetic field~\cite{Buividovich:2010tn}. This state could be a precursor of the superconducting phase with $ \partial_0 j_z \sim E_z$ which we discuss in this talk.}~\cite{ref:I}:
\beqn
\frac{\partial {\overline j}_z}{\partial t} = - \frac{2 e^3}{g^2_s} (B - B_c) E_z\,, 
\quad
\frac{\partial {\overline j}_{x,y}}{\partial t} = 0\,,
\label{eq:London:AS}
\eeqn
where ${\overline j}_i$, $i=x,y,z$ is the electric current averaged over one unit $\rho$-vortex cell. For simplicity, we used in Eqs.~\eq{eq:London:AS}  the averaged currents ${\overline j}_i$ instead of the original currents $j_i(x_\perp)$ because the value of the current depends nonlocally on the inhomogeneous order parameter $\rho = \rho(x_\perp)$, which would otherwise enter the right hand side of Eqs.~\eq{eq:London:AS}. The London-like equations~\eq{eq:London:AS} correspond to a linear response of the system in $E$, and they are valid in the domain
$E \ll B_c < B$ and $B - B_c \ll B_c$.

The transport equations~\eq{eq:London:AS} indicate that the vacuum state~\eq{eq:ud:cond} superconducts along the direction of the magnetic field, while in the transverse directions the electromagnetic superconductivity is absent.

One can rewrite Eqs.~\eq{eq:London:AS} in a rotationally-covariant form similarly to Eq.~\eq{eq:London},
\beqn
\partial_t \vec j+\vec \nabla q = - \frac{2 e^3}{g^2_s} \frac{|\vec B| - B_c}{(\vec B)^2} ({\vec B} \cdot {\vec E}) {\vec B}\,.
\label{eq:London:AS:3}
\eeqn

Introducing the four-current $j^\mu = (q,\vec j)$ one can generalize Eq.~\eq{eq:London:AS:3} to an explicitly Lorentz-covariant form:
\beqn
\partial_{[\mu,} j_{\nu]} =\kappa \frac{(F \cdot {\widetilde F})}{(F \cdot F)} {\widetilde F}_{\mu\nu}\,,
\eeqn
where the prefactor $\kappa$ is, in general, a scalar function of two Lorentz invariants $(F \cdot F) = F^{\mu\nu} F_{\mu\nu} \equiv 2 (\vec B^2 - \vec E^2)$ and 
$(F \cdot {\widetilde F}) = F^{\mu\nu} {\widetilde F}_{\mu\nu} \equiv 4 (\vec B \cdot \vec E)$ with 
${\widetilde F}_{\mu\nu} = \frac{1}{2} \epsilon_{\mu\nu\alpha\beta} F^{\alpha\beta}$.
In Eq.~\eq{eq:London:AS:3}  $\kappa = (e^3/g^2_s) (\sqrt{(F \cdot F)/2} - B_c)$ in the linear response approximation (with respect to the electric field $E$).
The cell-averaged curl of the {\emph {induced}} superconducting current is always zero, $\vec \nabla \times \vec j = 0$.

Finally, one can ask a provocative question.  All known superconductors expel weak external magnetic field exhibiting the Meissner effect. Moreover, if the magnetic field becomes strong enough the ordinary superconductivity is always destroyed. So why the magnetic field can penetrate the QCD superconductor without being suppressed by the superconducting state or without destroying the superconductivity itself? The answer is rather simple: these unexpected properties are due to the absence of the {\it transverse} superconductivity~\eq{eq:London:AS} in the condensed state~\eq{eq:ud:cond}. Indeed, the usual Meissner effect is caused by the magnetic-field-induced superconducting currents which circulate in the transverse (to the magnetic field) plane. If the magnetic field is so strong that the ordinary superconductor cannot support these large transverse currents (i.e., from the point of view of the energy balance), then the superconductor experiences a transition into a normal nonsuperconducting state. In the absence of the transverse superconductivity, the external magnetic field cannot be screened by the longitudinally superconducting QCD state and, moreover, this state cannot be destroyed by the strong magnetic field.

We hope that our idea will be checked in other models and in numerical simulations of lattice QCD. If true, it would be fascinating to observe how the empty space becomes an electromagnetic superconductor in a sufficiently strong external magnetic field.

\bibliographystyle{aipproc}

\end{document}